\documentclass[fleqn,twoside]{article}
\usepackage{espcrc2}

\usepackage{graphicx}
\usepackage[figuresright]{rotating}

\def\MZ{M_Z}

\def\dah{\Delta\alpha^{(5)}_{\rm had}}
\def\dahs{\Delta\alpha^{(5)}_{\rm had}(s)}
\def\dahz{\Delta\alpha^{(5)}_{\rm had}(\MZ^2)}
\def\dah0{\Delta\alpha^{(5)}_{\rm had}(-M_0^2)}
\def\dahm0{\Delta\alpha^{(5)}_{\rm had}(-M^2_0)}

\newcommand{\MSb}{$\overline{\mathrm{MS}}$ }
\newcommand{\epo}{\;\:.}

\newcommand{\noi}{\noindent}
\newcommand{\litem}{\noi \hspace*{3mm}$\bullet$~}
\newcommand{\bit}{\begin{itemize}}
\newcommand{\eit}{\end{itemize}}
\newcommand{\epm}{e^+e^-}
\newcommand{\mz}{M_Z^2}
\newcommand{\power}[1]{\times 10^{#1}}
\newcommand{\nn}{\nonumber}
\newcommand{\gv}{\mbox{GeV}}

\newcommand{\ba}{\begin{eqnarray*}}
\newcommand{\bea}{\begin{eqnarray}}
\newcommand{\be}{\begin{eqnarray*}}
\newcommand{\beq}{\begin{equation}}
\newcommand{\ea}{\end{eqnarray*}}
\newcommand{\eea}{\end{eqnarray}}
\newcommand{\ee}{\end{eqnarray*}}
\newcommand{\eeq}{\end{equation}}

\newcommand{\AmS}{{\protect\the\textfont2
  A\kern-.1667em\lower.5ex\hbox{M}\kern-.125emS}}
\newcommand{\bary}{\begin{array}}
\newcommand{\eary}{\end{array}}

\DeclareMathSymbol{\varPhi}{\mathalpha}{operators}{"08}
\DeclareMathSymbol{\varOmega}{\mathalpha}{operators}{"0A}
\DeclareMathSymbol{\varPsi}{\mathalpha}{operators}{"09}

\title{The running fine structure constant $\alpha(E)$ via the Adler function}

\author{F. Jegerlehner\address[IFJ]{H. Niewodniczanski Nuclear Physics
Institute PAN, 31-342 Krakow, Poland}%
\address[HUB]{Humboldt-Universit\"at zu Berlin,
Institut f\"ur Physik, Newtonstrasse 15, D-12489 Berlin, Germany}%
        \thanks{Permanent address.}%
\address[DEZ]{Deutsches Elekronen-Synchrotron DESY, Platanenalle 6,
D-15738 Zeuthen, Germany}%
        \thanks{Work supported in part by the EU grants
MTKD-CT2004-510126 in partnership with the CERN Physics Department and
with the TARI Program under contract RII3-CT-2004-506078.}%
                }
       
\begin{document}
\onecolumn{
\renewcommand{\thefootnote}{\fnsymbol{footnote}}
\setlength{\baselineskip}{0.52cm}
\thispagestyle{empty}
\begin{flushright} \begin{tabular}{c}
IFJPAN-IV-2008-3\\
HU-EP-08/22\\ 
DESY 08-078 \\
June 2008 \end{tabular}
\end{flushright}

\setcounter{page}{0}

\mbox{}
\vspace*{\fill}
\begin{center}
{\Large\bf 
The running fine structure constant $\alpha(E)$ via the Adler function}

\vspace{5em}
\large
F. Jegerlehner\footnote[1]{\noindent 
Work supported in part by the EU grants
MTKD-CT2004-510126 in partnership with the CERN Physics Department and
with the TARI Program under contract RII3-CT-2004-506078.}
\\
\vspace{5em}
\normalsize
{\it H. Niewodniczanski Nuclear Physics
Institute PAN, 31-342 Krakow, Poland}\\
{\it Humboldt-Universit\"at zu Berlin, Institut f\"ur Physik,\\
       Newtonstrasse 15, D-12489 Berlin, Germany\footnote[2]{\noindent
Permanent Address}}\\

        and

{\it Deutsches Elektronen-Synchrotron DESY,\\
       Platanenallee 6, D-15738 Zeuthen, Germany}\\
\end{center}
\vspace*{\fill}}
\newpage

\begin{abstract}
We present an up-to-date analysis for a precise determination of the
effective fine structure constant and discuss the prospects for future
improvements. We advocate to use a determination monitored by the Adler
function which allows us to exploit perturbative QCD in an optimal
well controlled way. Together with a long term program of hadronic
cross section measurements at energies up to a few GeV, a determination
of $\alpha(M_Z)$ at a precision comparable to the one of the $Z$ mass
$M_Z$ should be feasible. Presently $\alpha(E)$ at $E>1~\gv$ is the
least precisely known of the fundamental parameters of the SM. Since, in spite
of substantial progress due to new BaBar exclusive data, the region
1.4 to 2.4 GeV remains the most problematic one a major step in the
reduction of the uncertainties are expected from VEPP-2000~\cite{Eidelman:2006cn} and from a
possible ``high-energy'' option DAFNE-2 at
Frascati~\cite{Ambrosino:2006gka}. The up-to-date evaluation 
reads $\Delta\alpha^{(5)}_{\rm had}(M_Z^2) =  0.027515 \pm 0.000149$ or
$\alpha^{-1}(\mz)=128.957\pm0.020$.
\vspace{1pc}
\end{abstract}

\maketitle

\section{INTRODUCTION}
\begin{figure}[h]
\centering
\includegraphics[height=3.8cm]{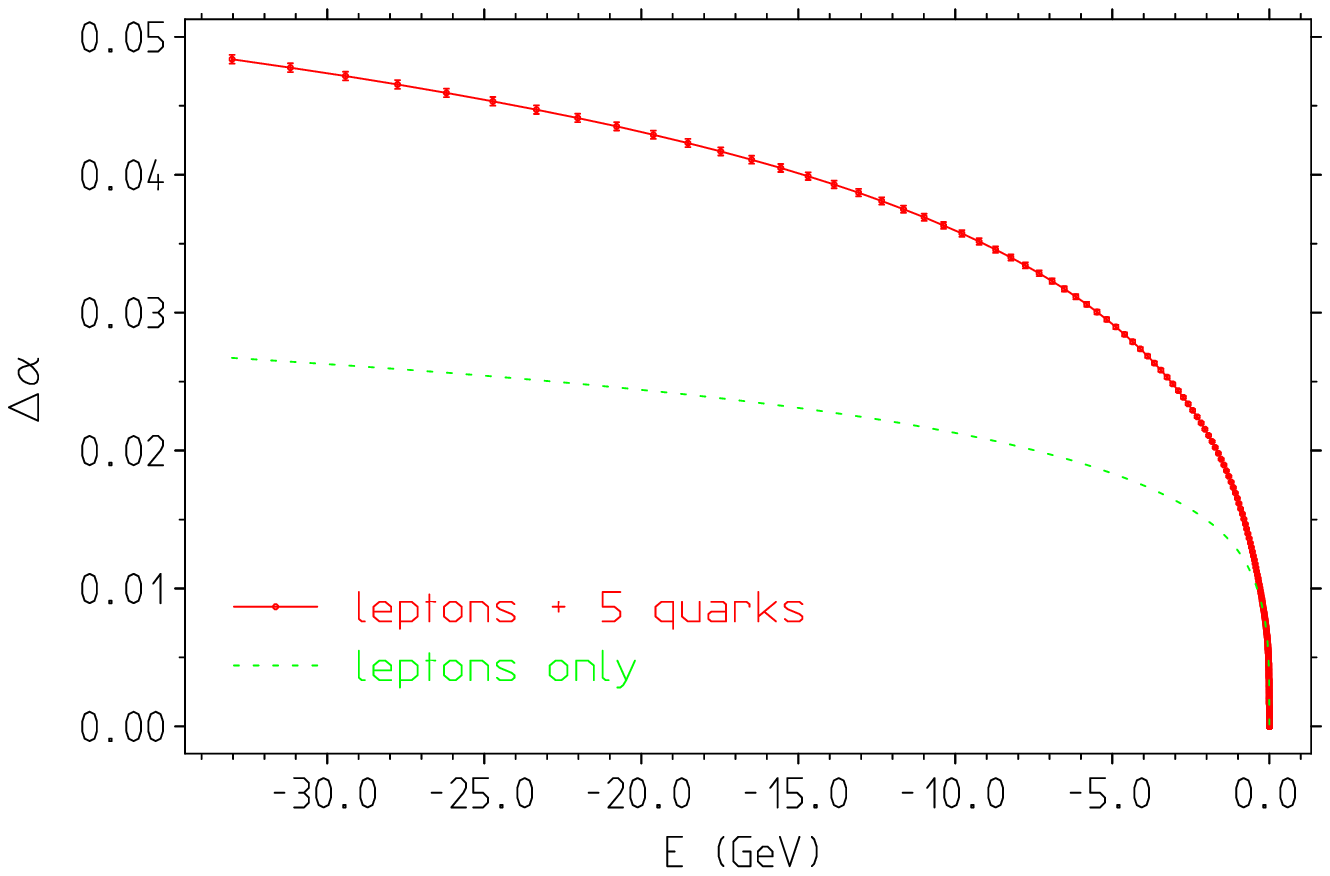}
\includegraphics[height=3.8cm]{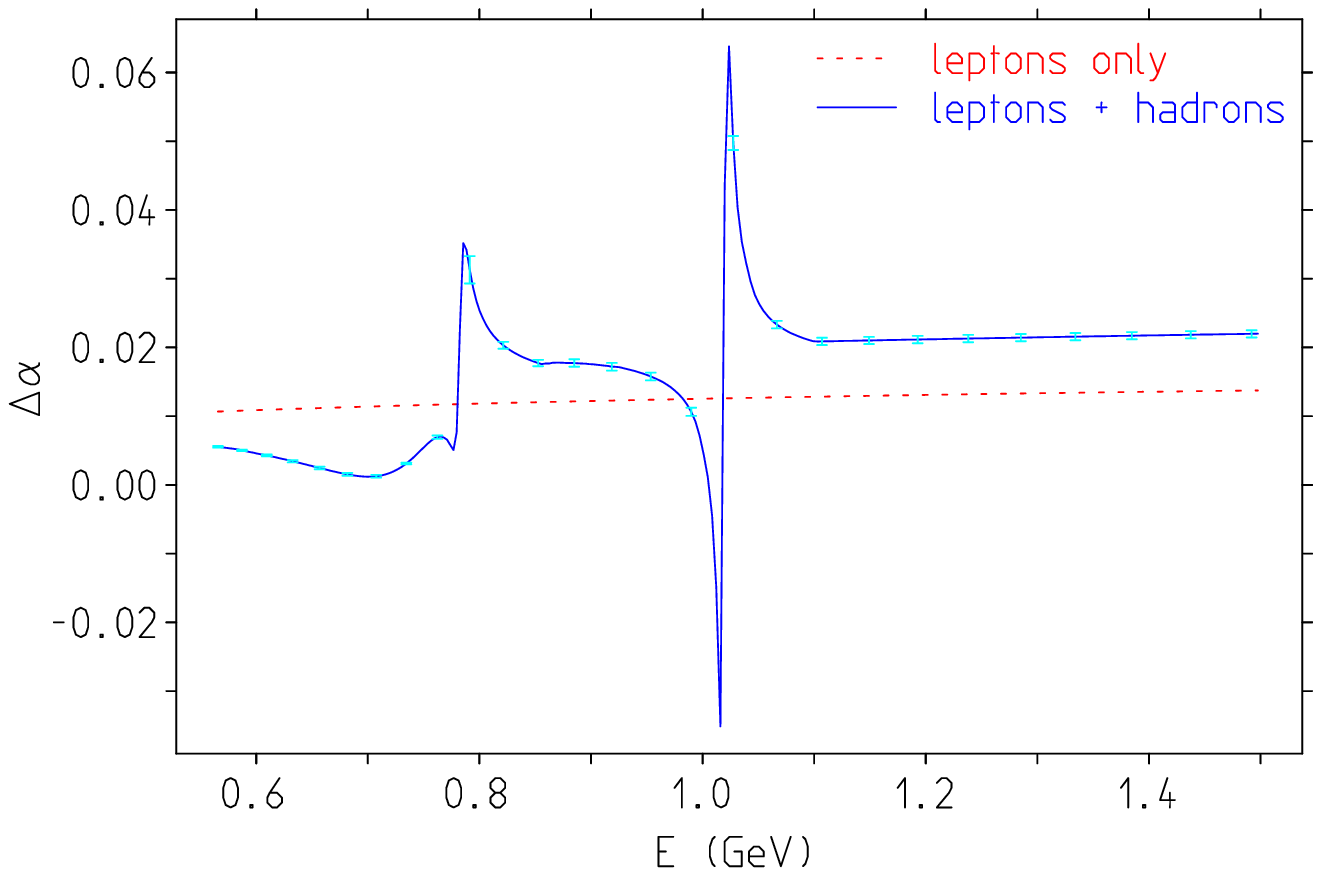}
\vspace*{-9mm}
\caption{The running of $\alpha$. The ``negative'' $E$ axis is
assigned to space-like momentum transfer. In the time-like region the
resonances lead to pronounced variations of the effective charge.}
\label{fig:runalp} 
\vspace*{-7mm}
\end{figure}
The accuracy of theoretical predictions of precision observables often
is limited as soon as low energy hadronic physics comes into play. In
fact, one of the main non-perturbative hadronic effect contributing to
many electroweak precision observables is the hadronic vacuum
polarization which affects the effective fine structure ``constant''
$\alpha(E)$. For precise SM predictions one thus needs to know the
running $\alpha$ very precisely. As $\alpha(E)$ is steeply
increasing at low $E$, substantial corrections show up at low scales
already. Furthermore, in the time--like region, non--perturbative
resonance effects make $\alpha(E)$ to be a complicated function, as
illustrated in Fig.~\ref{fig:runalp}.
\section{$\alpha(M_Z)$ IN PRECISION PHYSICS}
For SM predictions the most precisely known parameters $\alpha$,
$G_\mu$ and $M_Z$ are chosen as the basic input parameters. However,
for processes beyond the very low energy region not $\alpha$ itself
but $\alpha(E)$ plays the role of $\alpha$. This has dramatic
consequences for precision physics: the uncertainties of the hadronic
contributions to the effective $\alpha$ represent a major limitation
for electroweak precision physics. In fact $\alpha(E)$ above about
1GeV is a factor of 10 less well known than the next worse which is
the $Z$ mass $M_Z$. This is particularly important for a precise
investigation of $Z$ and $W$ gauge boson physics, for example.  The
present accuracies of the main SM parameters read $\delta \alpha /
\alpha \sim 3.7 \times 10^{-10}$, $\delta G_\mu / G_\mu \sim 8.6
\times 10^{-6}$, $\delta M_Z / M_Z \sim 2.4
\times 10^{-5}$, but $\delta \alpha(M_Z) / \alpha(M_Z) \sim 1.1 \div 2.6
\times 10^{-4}$ [$\delta \alpha(E) / \alpha(E) \sim 4.5 
\times 10^{-5}$ at $E=1~\gv$ space-like].  Thus at present we loose a factor $10^{5}$ in
precision in the replacement $\alpha
\to \alpha(M_Z)$. For precision physics at the ILC one would require
$\alpha(M_Z)$ to be determined as precise as $M_Z$~\cite{fjeger01}, typically, which
would require an improvement by a factor about 10 to obtain
\bea 
\bary{ccccc}
{\frac{\delta \alpha(M_Z)}{\alpha(M_Z)}} &{\sim}& {2.5 }& {
\times}& {10^{-5}} \epo
\eary
\label{goal}
\eea
At present, an important example is the LEP/SLD measurement of
$\sin^2 \Theta_{\rm eff}=(1-g_{Vl}/g_{Al})/4 =0.23148 \pm 0.00017$
from which the Higgs mass bound depends most sensitively. An uncertainty
of $\delta \Delta \alpha(M_Z)=0.00036$ leads to an error $\delta
\sin^2 \Theta_{\rm eff}=0.00013$ in the prediction of $\sin^2
\Theta_{\rm eff}$,  and
affects the Higgs mass bound, precision tests and new physics searches.
Note that once $m_H$ will have been measured by the LHC
$\sin^2 \Theta_{\rm eff}$ will be an excellent monitor for new physics!
This will be particularly important once the top mass $m_t$ will have
been determined with higher accuracy.

One also should keep in mind that for calculations of perturbative QCD
contributions precise QCD parameters $\alpha_s, \; m_c, \; m_b,\;
m_t$ are mandatory.

\section{UPDATED EVALUATION OF $\alpha(M_Z)$}
Since my last major update in August 2006 a number of new results
mainly from BaBar~\cite{BaBar07} were published. In fact a series of new
channels have been measured in a range which covers the problematic
region between 1.4 and 2.4 GeV. This means that we have almost
completely new data for the exclusive measurements in this region. In
contrast the inclusive measurements date back to the early
1980's. Important new cross--section
measurements  were also presented by KLOE at this meeting~\cite{FNguyen} (see
also status reports from CMD-2/SND, BaBar, Belle, CLEO and BES at this
meeting).  The standard evaluation of the non-perturbative hadronic
contributions in terms of measured cross-sections $\sigma(e^+e^- \to
{\rm hadrons})$ is based on the dispersion integral:
\bea
\dahs &=& - \frac{\alpha s}{3\pi}\;\bigg(\;\;\;
{\rm \footnotesize P}\!\!\!\!\!\!\!\!  \int\limits_{4m_\pi^2}^{E^2_{\rm cut}} ds'
\frac{{R_\gamma(s')}}{s'(s'-s)} 
\bigg)\,,
\label{DIdirect}
\eea
where the $\epm$--data are encoded in $R_\gamma(s) \equiv
{\sigma^{(0)}(e^+e^- \rightarrow \gamma^*
\rightarrow {\rm hadrons})}/{\frac{4\pi \alpha^2}{3s}}$.

The evaluation of the integral at $M_Z=$ 91.19 GeV is performed by
using $R(s)$ data up to $\sqrt{s}=E_{\rm cut}=5$ GeV
and for the $\Upsilon$ resonances region between 9.6 and 13 GeV.
Perturbative QCD is applied from 5.0 to 9.6 GeV and for the high energy tail above 13 GeV.
The result is
\bea
\Delta \alpha _{\rm hadrons}^{(5)}(\mz)  & = & 0.027594 \pm 0.000219\nn \\
\alpha^{-1}(\mz)& = & 128.946 \pm 0.030\;.
\eea
Note that BaBar exclusive radiative return measurements in this
evaluation play an essential role up to 2 GeV [lower end of BES
inclusive measurement]. In the problematic region from 1.4 to 2 GeV
the exclusive measurements actually dominate in comparison to the much
older inclusive measurements from Frascati [MEA, $\gamma \gamma$2,
M3N, B$\overline{\mathrm{B}}$]. More detail are given in Table~\ref{tab:resdirect}.
 
\begin{table*}[t]
\begin{center}
\caption{Contributions and uncertainties for 
$\dahz^{\mathrm{data}}\power{4}$.}
\label{tab:resdirect}
\begin{tabular}{ccrrr}
\hline
 final state &  range (GeV) & result~~(stat)~~(syst)~~[tot]~~ & rel~ & abs~ \\
\hline
   $\rho    $   & (0.28, 0.81) &     25.95 ( 0.09) ( 0.14)[ 0.17]&  0.6\% &  0.6\% \\
   $\omega  $   & (0.42, 0.81) &      2.91 ( 0.03) ( 0.08)[ 0.09]&  3.0\% &  0.2\% \\
   $\phi    $   & (1.00, 1.04) &      4.42 ( 0.06) ( 0.10)[ 0.12]&  2.7\% &  0.3\% \\
   $J/\psi  $   &             &     11.14 ( 0.53) ( 0.58)[ 0.79]&  7.1\% & 12.9\% \\
   $\Upsilon$   &             &      1.18 ( 0.05) ( 0.06)[ 0.08]&  6.9\% &  0.1\% \\
     had        & (0.81, 1.40) &     13.21 ( 0.04) ( 0.35)[ 0.35]&  2.7\% &  2.6\% \\
     had        & (1.40, 2.00) &     11.34 ( 0.07) ( 1.26)[ 1.26]& 11.2\% & 33.3\% \\
     had        & (2.00, 3.10) &     15.73 ( 0.11) ( 0.87)[ 0.88]&  5.6\% & 16.2\% \\
     had        & (3.10, 3.60) &      5.26 ( 0.11) ( 0.10)[ 0.15]&  2.8\% &  0.5\% \\
     had        & (3.60, 9.46) &     50.58 ( 0.11) ( 0.24)[ 0.26]&  0.5\% &  1.5\% \\
     had        & (9.46,13.00) &     18.52 ( 0.25) ( 1.21)[ 1.23]&  6.7\% & 31.8\% \\
    pQCD        & (13.0,$\infty$) &    115.71 ( 0.00) ( 0.06)[ 0.06]&  0.0\% &  0.1\% \\
 \hline
    data        & (0.28,13.00) &    160.23 ( 0.63) ( 2.10)[ 2.19]&  1.4\% &  0.0\% \\
    total       &             &    275.94 ( 0.63) ( 2.10)[ 2.19]&  0.8\% &100.0\% \\
 \hline
 \end{tabular}
\end{center}
\end{table*}

\section{TESTING NON--PERTURBATIVE HADRONIC EFFECTS VIA THE ADLER FUNCTION}
\begin{figure}[h]
\centering
\includegraphics[height=4.6cm]{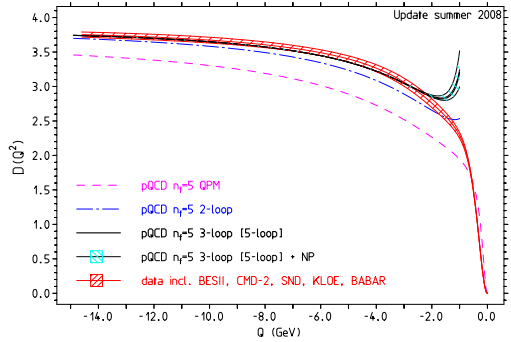}
\vspace*{-6mm}
\caption{The ``experimental'' non-perturbative Adler--function versus theory (pQCD + NP).
The error includes statistical + systematic here (in contrast to most
$R$-plots showing statistical errors only!). ``[5-loop]'' indicates
that 4- and 5-loop contribution in the massless limit are taken into account. For
more details see Ref.~\cite{EJKV98}.}
\label{fig:adler}
\end{figure}

The non-perturbative Adler function related to the photon vacuum polarization can be
calculated in terms of experimental $\epm$ annihilation data by the
dispersion integral
\bea
D(Q^2) &= &Q^2\:\left(\int_{4 m_{\pi}^2}^{E^2_{\rm cut}}
\frac{  R^{\rm data}(s)}{(s+Q^2)^2}\,ds\right.
\nn \\ &&~~~~~+ \!\!\left.
\int_{E^2_{\rm cut}}^{\infty}\frac{  R^{\rm pQCD}(s)}{(s+Q^2)^2}\,ds\:\right)\;.
\eea
Here $Q^2=-q^2$ is the squared Euclidean momentum transfer and 
$s$ the center of mass energy squared for hadron production in
$\epm$--annihilation.
Formally the Adler function is defined as the derivative of the 
shift in the fine structure constant
\ba
\frac{D(Q^2)}{Q^2}=(12\pi^2)\,\frac{d\Pi'_{\gamma}\,(q^2)}{dq^2}
=-\frac{3\pi}{\alpha}\frac{d}{dq^2}\Delta \alpha_{\mathrm{had}}(q^2)\;,
\ea
evaluated in the Euclidean at $Q^2=-q^2$. $\Pi'_{\gamma}\,(q^2)$ is the
photon vacuum polarization amplitude defined by
\bea 
\Pi^{\gamma}_{\mu \nu}(q) &=& i
\int d^4 x e^{iqx} <0|T J^\gamma_\mu\; (x)\; J^\gamma_\nu\;(0)\;|0>
\nn \\
&=& -\left(q^2\, g_{\mu \nu}- q_\mu q_\nu\right)\;
\Pi'_{\gamma}\;(q^2)\;.
\eea 
The perturbative result is given in~\cite{EJKV98}. Crucial for this
prediction are known full massive QCD
results~\cite{Chetyrkin:1996cf,Chetyrkin:1997qi,JT98}.  Note that the
main $Q^2$ dependence of $D(Q^2)$ is due to the quark masses $m_c$ and
$m_b$. Without mass effects, up to small effects from the running of
$\alpha_s$, $D(Q^2)=3\sum_f Q_f^2\,(1+O(\alpha_s))$ is a constant
depending on the number of active flavors. We also include 
the 4--loop~\cite{Gorishnii:1990vf,Surguladze:1990tg}
and 5--loop~\cite{Baikov:2001aa} contributions in the high energy limit (massless
approximation)
$$D(Q^2)\simeq 3\sum_f Q_f^2\,\left(1+a+d_2 a^2+d_3 a^3 +d_4 a^4\right)$$
with $a=\alpha_s (Q^2)/\pi$, 
$d_2=1.9857-0.1153\, n_f$, 
$d_3=18.2428-4.2159\, n_f + 0.0862\, n_f^2 -1.2395\, (\sum Q_f)^2/(3
\sum Q_f^2)$ and $d_4=-0.010\,n_f^3\,+\,1.88\,n_f^2\,-\,34.4\,n_f\,+\,135.8$.
The corresponding formula for $R(s)$ only differs at the 4--loop and
5--loop level due to the effect from the analytic continuation from
the Euclidean to the Minkowski region which yields
$r_3^{R}=d_3-\pi^2\beta_0^2 \frac{d_1}{3}$ with
$\beta_0=(11-2/3\, n_f)/4$, $d_1=1$ and
$r_4^{R}=d_4-\pi^2\beta_0^2
\left(d_2+\frac{5\beta_1}{6\beta_0}\,d_1\right)$ with
$\beta_1=(102-38/3\,n_f)/16$. Numerically the 4--loop term
proportional to $d_3$ amounts to $-0.0036$\% at 100 GeV and increases
to about $0.32$\% at 2.5 GeV. The higher order massless results only
improve the perturbative high energy tail (see Fig.~\ref{fig:adler}).
Towards low $Q^2$ we also approach the Landau pole of $\alpha_s(Q^2)$,
present typically in \MSb type schemes, and pQCD ceases to ``converge''. 

\section{$\Delta \alpha^{\rm had}$ VIA THE ADLER FUNCTION}
Figure~\ref{fig:adler} provides convincing evidence that pQCD works well to predict $D(Q^2)$ down to
$Q\sim M_0= 2.5\, \gv $. This may be used to calculate 
\ba
\Delta \alpha_{\mathrm{had}}(-Q^2) \sim \frac{\alpha}{3\pi} \int
dQ^{'2} \frac{D(Q^{'2})}{Q^{'2}}
\ea
and we may write
\bea
&&\Delta\alpha^{(5)}_{\rm had}(M_Z^2) = \Delta\alpha^{(5)}_{\rm
had}(-M_0^2)^{\mathrm{data}} \nn \\ \hspace*{-1.6cm}&&+\,
\left[\Delta\alpha^{(5)}_{\rm had}(-M_Z^2) -\Delta\alpha^{(5)}_{\rm
had}(-M_0^2)\right]^{\mathrm{pQCD}} \nn \\ \hspace*{-1.6cm}&&+\,
\left[\Delta\alpha^{(5)}_{\rm had}(M_Z^2) -\Delta\alpha^{(5)}_{\rm
had}(-M_Z^2)\right]^{\mathrm{pQCD}}
\eea
and obtain, for $M_0=2.5\, \gv$
\bea
\Delta\alpha^{(5)}_{\rm had}(-M_0^2)^{\mathrm{data}} &=& 0.007354 \pm
0.000107\\
\Delta\alpha^{(5)}_{\rm had}(-M_Z^2) &=&  0.027477 \pm 0.000149\nn \\
\Delta\alpha^{(5)}_{\rm had}(M_Z^2) &=&  0.027515 \pm 0.000149 
\label{alpAdler}
\eea         
where a tiny shift of $+0.000008$ results from the 5-loop contribution.
An error $\pm 0.000103$ added in quadrature comes form the perturbative part.
For the perturbative calculation of $\left[\Delta\alpha^{(5)}_{\rm had}(-M_Z^2) -\Delta\alpha^{(5)}_{\rm
had}(-M_0^2)\right]^{\mathrm{pQCD}}$ we use the QCD parameters: $\alpha_s(M_Z)=0.1189(20)$,\\
\centerline{$m_c(m_c)=1.286(13)~[M_c=1.666(17)]~\gv\,,$}\\
\centerline{$m_b(m_c)=4.164(25)~[M_b=4.800(29)]~\gv\,,$}\\
based on a complete 3--loop massive QCD analysis~\cite{Kuhn:2007vp,BCS06}
(see contributions by K\"uhn and Sturm). Note that due to a dramatic
improvement in the determination of the quark masses $m_c$ and $m_b$,
for the first time the pQCD error included in (\ref{alpAdler}) is
smaller than the one from the data which also has been improved
substantially. A very important long term project here is the lattice
determinations of the basic QCD parameters~\cite{Heitger}.
  
\begin{table*}[t]
\begin{center}
\caption{Contributions and uncertainties for
$\dahm0^{\mathrm{data}}\power{4}$ ($M_0=2.5$ GeV).}
\label{tab:resAdler}
\begin{tabular}{ccrrr}
\hline
 final state &  range (GeV) & result~~(stat)~~(syst)~~[tot]~~ & rel~ & abs~ \\
\hline
   $\rho    $   & (0.28, 0.81) &     24.06 ( 0.09) ( 0.13)[ 0.16]&  0.6\% &  2.1\% \\
   $\omega  $   & (0.42, 0.81) &      2.65 ( 0.03) ( 0.07)[ 0.08]&  3.0\% &  0.5\% \\
   $\phi    $   & (1.00, 1.04) &      3.79 ( 0.05) ( 0.09)[ 0.10]&  2.7\% &  0.9\% \\
   $J/\psi  $   &             &      3.95 ( 0.19) ( 0.18)[ 0.26]&  6.6\% &  5.9\% \\
   $\Upsilon$   &             &      0.07 ( 0.00) ( 0.00)[ 0.00]&  6.7\% &  0.0\% \\
     had        & (0.81, 1.40) &     11.33 ( 0.03) ( 0.29)[ 0.29]&  2.6\% &  7.3\% \\
     had        & (1.40, 2.00) &      7.81 ( 0.05) ( 0.87)[ 0.87]& 11.2\% & 65.8\% \\
     had        & (2.00, 3.10) &      7.91 ( 0.05) ( 0.44)[ 0.44]&  5.6\% & 16.7\% \\
     had        & (3.10, 3.60) &      1.88 ( 0.04) ( 0.04)[ 0.05]&  2.8\% &  0.2\% \\
     had        & (3.60, 9.46) &      8.11 ( 0.02) ( 0.05)[ 0.05]&  0.6\% &  0.2\% \\
     had        & (9.46,13.00) &      0.89 ( 0.01) ( 0.06)[ 0.06]&  6.6\% &  0.3\% \\
    pQCD        & (13.0,$\infty$) &      1.09 ( 0.00) ( 0.00)[ 0.00]&  0.1\% &  0.0\% \\
 \hline
    data        & (0.28,13.00) &     72.45 ( 0.23) ( 1.05)[ 1.08]&  1.5\% &  0.0\% \\
    total       &             &     73.54 ( 0.23) ( 1.05)[ 1.08]&  1.5\% &100.0\% \\
\hline
\end{tabular}
\end{center}
\end{table*}

\begin{figure}[t]
\centering
\includegraphics[height=2.7cm]{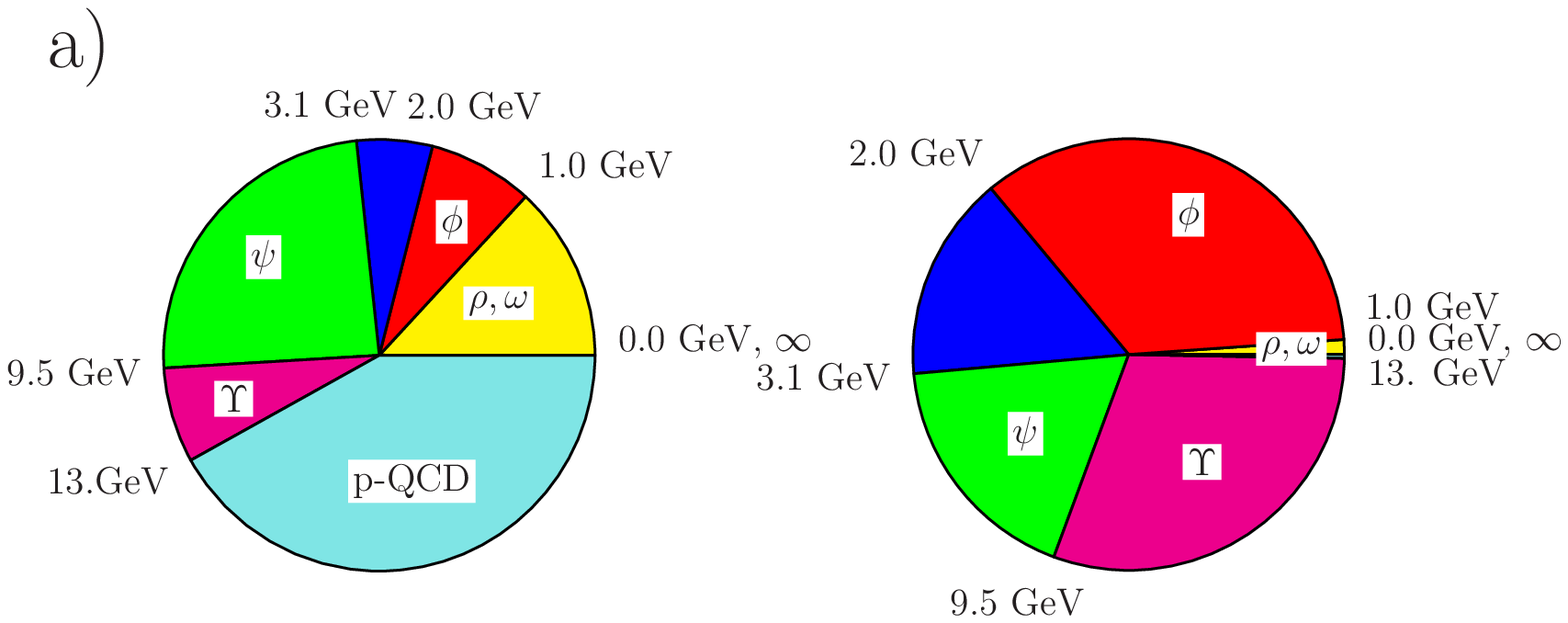}
\includegraphics[height=2.7cm]{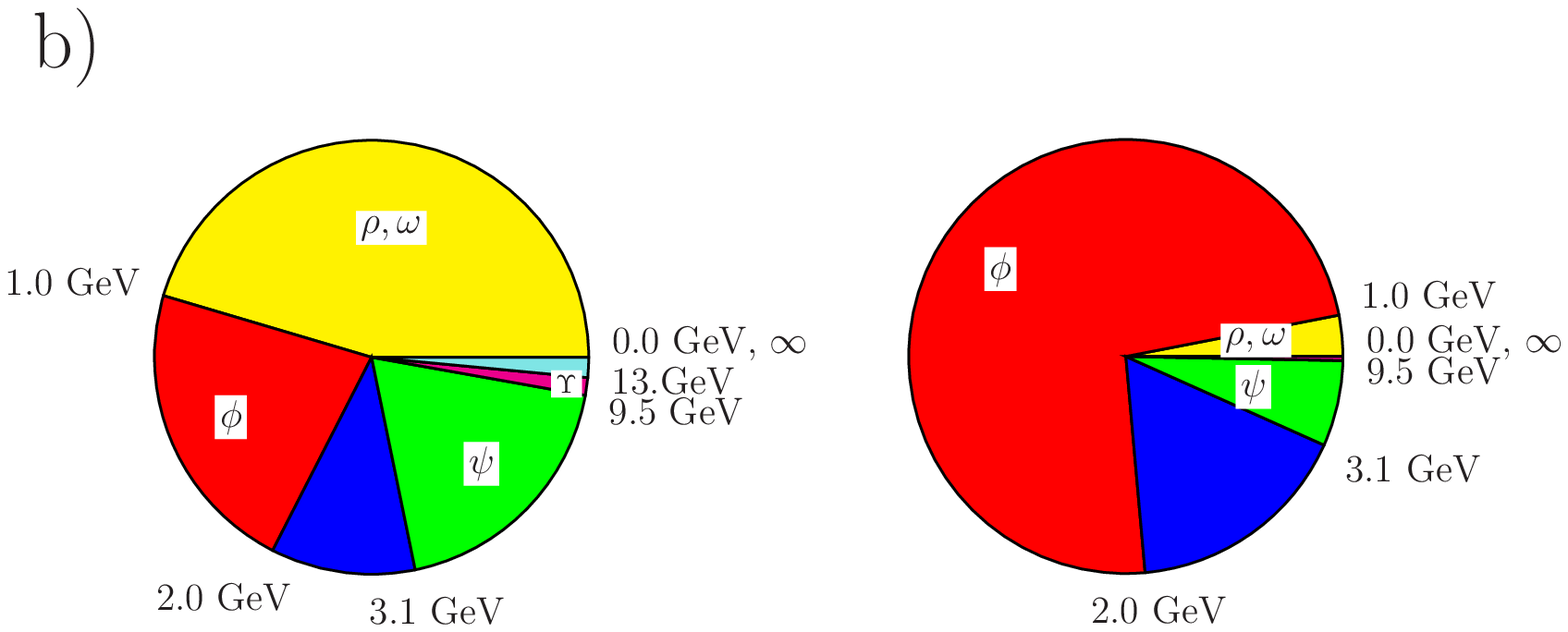}
\vspace*{-12mm}
\caption{Present distribution of contributions [left] and errors$^2$
[right] for
a) $\Delta \alpha _{\rm hadrons}^{(5)}(\mz)$;
b) $\dahm0^{\mathrm{data}}$ ($M_0=2.5$ GeV);
both obtained by direct integration of (\ref{DIdirect}).}
\label{fig:distrib} 
\vspace*{-6mm}
\end{figure}

Mandatory pQCD improvements required are:\\
$\bullet$ 4--loop massive pQCD calculation of Adler function; 
required are a number of terms in the low and high momentum series expansions which
allow for the appropriate Pad\'e improvements
[essentially equivalent to a massive 4--loop calculation of $R(s)$];\\
$\bullet$ $m_c$ improvement by sum rule and/or lattice QCD evaluations;\\
$\bullet$ improved $\alpha_s$ in low $Q^2$ region.  

Renormalization schemes which exhibit a Landau pole, like the \MSb scheme, evidently fail in
parametrizing the low energy tail of the Adler function. Therefore
modeling the Adler function at low $Q^2$ 
by testable models may be useful, such as ``analytized'' $\alpha_s$~\cite{Shirkov06} and
the instanton liquid model~\cite{Dorokhov04} or -- others.

\begin{figure}[t]
\centering
\includegraphics[height=7.7cm]{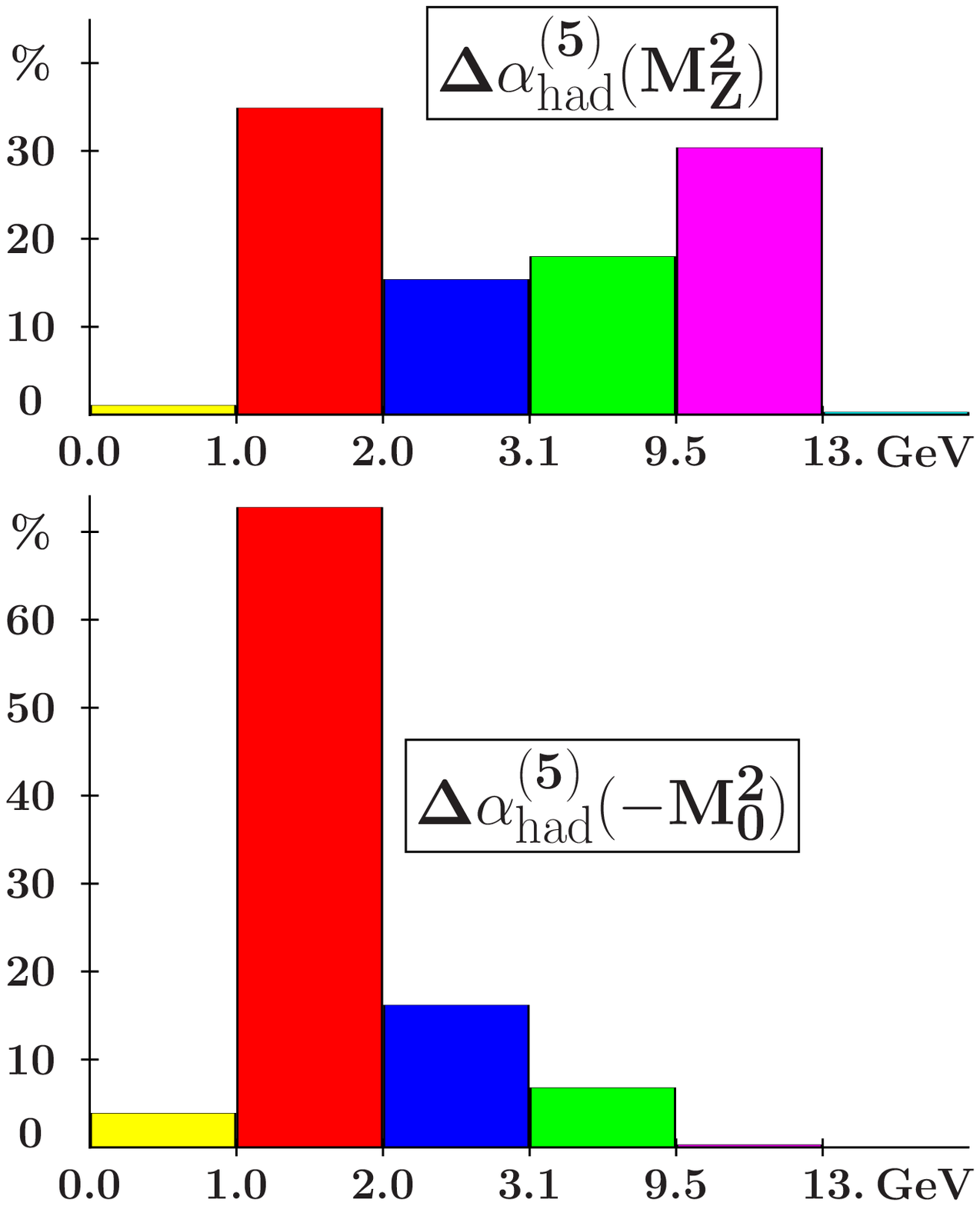}
\vspace*{-6mm}
\caption{Present error profiles for $\dahz$ and $\dah0$.}
\label{fig:errorprof} 
\vspace*{-8mm}
\end{figure}

The contribution and error profiles of $\dahz$ and $\dah0$, are shown
in Fig.~\ref{fig:distrib}. Fig.~\ref{fig:errorprof} illustrates where
more precise measurements are particularly important. For the Adler
function approach in particular low energy machines below 2.5 GeV most
successfully can contribute to improve the precise determination of
$\alpha(E)$. At the same time machines in this regime substantially
contribute to further reduce the error of the leading hadronic
contribution to the muon $g-2$.

\section{POSSIBLE IMPROVEMENT BY VEPP-2000 and DAFNE-2}

Next generation precision physics experiments, not only the ones
possible at an ILC, in many cases require a more precise determination
of $\alpha(E)$. A reasonable goal could be an improvement by about a
factor 10 in accuracy which would match the precision of the $Z$ mass.
The options are

\litem the standard approach by direct integration of the $\epm$--data: in this case
58\% of the contribution is obtained from the data and 42\% from
pQCD. My analysis yields
$\Delta \alpha_{{\rm had}}^{(5)\:\mathrm{data}} \times 10^4=160.12\pm 2.24$ (1.4\%)
and thus increasing the overall accuracy to 1\% would yield an
uncertainty $\pm 1.63$. However, for independent measurements in
ranges as used in the Tab.~\ref{tab:resdirect} a
1\% accuracy for each region and errors including systematic ones added in
quadrature would yield  $\pm 0.85$. 
The improvement on the data  ([2.24] vs. [0.85]) thus would yield an
improvement factor of 2.6. The pQCD part in this case is
$\Delta \alpha_{{\rm had}}^{(5)\:\mathrm{pQCD}} 
\times 10^4=115.71\pm 0.06$ (0.05\%) and for the theory part this means
that no improvement would be needed.\\
\litem With the ``Adler function approach'' we get 26\% of the contribution
from data and 74\% from pQCD. Here $\Delta \alpha_{{\rm had}}^{(5)\:\mathrm{data}} 
\times 10^4=72.35\pm 1.10$ (1.5\%) and a
1\% overall accuracy  would mean an uncertainty $\pm 0.74$. Again, a
subdivision of ranges as used in
Tab.~\ref{tab:resAdler} and assuming that a
1\% accuracy can be reached for each region and  adding up errors
in quadrature in this case would lead to a precision of $\pm 0.40$. 
The improvement from the data ([1.10] vs. [0.40]) again yields a
similar improvement factor of 2.7. If we compare the standard
approach of direct integration with the Adler function controlled approach
([2.24] vs. [0.40]) we have an  improvement factor 5.6. However, now
a much larger fraction $\Delta \alpha_{{\rm had}}^{(5)\:\mathrm{pQCD}} 
\times 10^4=201.15\pm 1.03$ is coming from pQCD and an improvement of
the QCD prediction is mandatory in order to profit in an optimal way
from the improvement on the data. A factor 3 to 5 at least should be possible
in a long term effort on higher order effects and more importantly on QCD parameters. 
An accuracy of about $\pm 0.20$ would be a high goal.

Our study shows that the requirement Eq.~(\ref{goal}) could be
achieved by\\
$\bullet$ pinning down experimental errors to the 1\% level in all
non-perturbative regions up to 10 GeV\\
$\bullet$ safely use pQCD in the Euclidean region monitored by the Adler function\\
$\bullet$ improve on pQCD and QCD parameters.\\
In any case as we see from Fig.~\ref{fig:errorprof} by far the largest
improvement factor will come from precise cross--section measurements in the 
region from 1.4 to 2.4 GeV. A unique challenge and chance for VEPP-2000
and DAFNE-2.

{\bf Acknowledgments:} I am very grateful to the organizers of the
Frascati Int. Workshop ``$\epm$ Collisions from $\varPhi$ to $\varPsi$''
for the kind invitation and for support. It was a very interesting and 
inspiring event. I also thank S. Jadach and his team from the
Niewodniczanski Nuclear Physics Institute at Krakow for the kind
hospitality extended to me.


\begin{thebibliography}{9}

\bibitem{Eidelman:2006cn}
B. Khazin, these proceedings; 
 S.~Eidelman,
  Nucl.\ Phys.\ Proc.\ Suppl.\  {\bf 162} (2006) 323.

\bibitem{Ambrosino:2006gka}
P.~Raimondi, these proceedings;
  G.~Venanzoni,
  Acta Phys.\ Polon.\  B {\bf 38} (2007) 3421;
  F.~Ambrosino et al.,
  Eur.\ Phys.\ J.\  C {\bf 50} (2007) 729.

\bibitem{fjeger01}
  F.~Jegerlehner,\textit{The effective fine structure constant at TESLA energies},
  hep-ph/0105283.

\bibitem{BaBar07}
B.~Aubert et al. [BABAR Collab.],
Phys.\ Rev.\  D {\bf 76} (2007) 012008;
Phys.\ Rev.\  D {\bf 76} (2007) 092005;
Phys.\ Rev.\  D {\bf 76} (2007) 092006;
arXiv:0710.4451 [hep-ex].

\bibitem{FNguyen}
F.~Nguyen [for the KLOE Collaboration],
  arXiv:0807.1612 [hep-ex].

\bibitem{FJ06}
F.~Jegerlehner,
Nucl.\ Phys.\ Proc.\ Suppl.\  {\bf 162} (2006) 22.

\bibitem{EJKV98}
S.~Eidelman, F.~Jegerlehner, A.~L.~Kataev, O.~Veretin,
Phys.\ Lett.\ B {\bf 454} (1999) 369.

\bibitem{FJ98} F. Jegerlehner, In: {\it Radiative Corrections},
ed by J.~Sol\`a (World Scientific, Singapore 1999) pp 75--89.

\bibitem{EJ95}
S.~Eidelman, F.~Jegerlehner,
Z.\ Phys.\ C {\bf 67} (1995) 585;
F.~Jegerlehner,
Nucl.\ Phys.\ (Proc.\ Suppl.) C  {\bf 51} (1996) 131;
%
J.\ Phys.\ G {\bf 29} (2003) 101;
Nucl.\ Phys.\ Proc.\ Suppl.\  {\bf 126} (2004) 325.

\bibitem{Chetyrkin:1996cf}
  K.~G.~Chetyrkin, J.~H.~K\"uhn, M.~Steinhauser,
  Nucl.\ Phys.\  B {\bf 482} (1996) 213;
  Nucl.\ Phys.\  B {\bf 505} (1997) 40.

\bibitem{Chetyrkin:1997qi}
  K.~G.~Chetyrkin, R.~Harlander, J.~H.~K\"uhn, M.~Steinhauser,
  Nucl.\ Phys.\  B {\bf 503} (1997) 339.

\bibitem{JT98}
  F.~Jegerlehner, O.~V.~Tarasov,
  Nucl.\ Phys.\  B {\bf 549} (1999) 481.

\bibitem{Gorishnii:1990vf}
  S.~G.~Gorishnii, A.~L.~Kataev, S.~A.~Larin,
  Phys.\ Lett.\  B {\bf 259} (1991) 144.

\bibitem{Surguladze:1990tg}
  L.~R.~Surguladze, M.~A.~Samuel,
  Phys.\ Rev.\ Lett.\  {\bf 66} (1991) 560
  [Erratum-ibid.\  {\bf 66} (1991) 2416].

\bibitem{Baikov:2001aa}
J.~H.~K\"uhn, these proceedings;
  P.~A.~Baikov, K.~G.~Chetyrkin, J.~H.~K\"uhn,
  Phys.\ Rev.\ Lett.\  {\bf 88} (2002) 012001;
%
  Phys.\ Rev.\  D {\bf 67} (2003) 074026;
%
  Phys.\ Lett.\  B {\bf 559} (2003) 245;
%
  arXiv:0801.1821 [hep-ph].

\bibitem{Kuhn:2007vp}
C.~Sturm, these preceedings;
  J.~H.~K\"uhn, M.~Steinhauser, C.~Sturm,
  Nucl.\ Phys.\  B {\bf 778} (2007) 192.
%
%
%
%
\bibitem{BCS06}
  R.~Boughezal, M.~Czakon, T.~Schutzmeier,
  Phys.\ Rev.\  D {\bf 74} (2006) 074006.
\bibitem{Heitger}
J.~Heitger, these proceedings;
  M.~Della Morte, N.~Garron, M.~Papinutto, R.~Sommer,
  JHEP {\bf 0701} (2007) 007;
  G.~M.~de Divitiis, M.~Guagnelli, R.~Petronzio, N.~Tantalo, F.~Palombi,
  Nucl.\ Phys.\  B {\bf 675} (2003) 309;
  J.~Rolf, S.~Sint  [ALPHA Collab.],
  JHEP {\bf 0212} (2002) 007.

\bibitem{Shirkov06}
  D.~V.~Shirkov, I.~L.~Solovtsov,
  Theor.\ Math.\ Phys.\  {\bf 150} (2007) 132.

\bibitem{Dorokhov04}
  A.~E.~Dorokhov,
  Phys.\ Rev.\  D {\bf 70} (2004) 094011.

\end{thebibliography}
\end{document}